**Antiferromagnetism-driven two-dimensional topological nodal-point superconductivity**

Maciej Bazarnik[1,2,*,§], Roberto Lo Conte[1,*,§], Eric Mascot[1,3,*,§], Dirk K. Morr[4], Kirsten von Bergmann[1], Roland Wiesendanger[1]

[1]*Department of Physics, University of Hamburg, D-20355 Hamburg, Germany*
[2]*Institute of Physics, Poznan University of Technology, Piotrowo 3, 60-965 Poznan, Poland*
[3]*School of Physics, University of Melbourne, Parkville, VIC 3010, Australia*
[4]*Department of Physics, University of Illinois at Chicago, Chicago, IL 60607*

**Abstract**

Magnet/superconductor hybrids (MSHs) hold the promise to host emergent topological superconducting phases [1, 2, 3, 4, 5]. Both one-dimensional (1D) [6, 7, 8] and two-dimensional (2D) [9, 10, 11, 12] magnetic systems in proximity to *s*-wave superconductors have shown evidence of gapped topological superconductivity with zero-energy end states and chiral edge modes. Recently, it was proposed [13] that the bulk transition-metal dichalcogenide 4Hb-TaS$_2$ is a gapless topological nodal-point superconductor (TNPSC) [14]. However, there has been no experimental realization of a TNPSC in a MSH system yet. Here we present the discovery of TNPSC in antiferromagnetic (AFM) monolayers on top of an *s*-wave superconductor. Our calculations show that the topological phase is driven by the AFM order, resulting in the emergence of a gapless time-reversal invariant topological superconducting state [15]. Using low-temperature scanning tunneling microscopy we observe a low-energy edge mode, which separates the topological phase from the trivial one, at the boundaries of antiferromagnetic islands. As predicted by the calculations, we find that the relative spectral weight of the edge mode depends on the edge's atomic configuration. Our results establish the combination of antiferromagnetism and superconductivity as a novel route to design 2D topological quantum phases.

In the last decade different material platforms have been proposed for the establishment of topological superconductivity. Among those are the semiconductor/superconductor [16], the topological insulator/superconductor [17], and the magnet/superconductor [1, 2, 3, 4, 5] platforms. While the first two platforms require a magnetic field for the stabilization of a topological superconducting phase, which is difficult to reconcile with miniaturization requirements of microelectronics, this issue is circumvented in the latter platform, the MSHs, where the presence of a magnet within the system can provide the required time-reversal symmetry breaking. So far, attention has been mainly focused on using ferromagnetic (FM) components in 1D and 2D MSHs [6, 7, 8, 9, 10, 11], which are understood as gapped topological superconductors described by a non-zero Chern number [18, 19]. However, hybrid systems with AFM order have not yet been considered. Another class of topological superconductivity, the gapless TNPSC phase, has been discussed extensively [20, 21, 22, 23, 24, 25, 26], and recently its observation was reported for the transition metal dichalcogenide $TaS_2$ [13], in the form of low energy modes at step edges. However, for MSHs no evidence of TNPSC has been reported so far.

In this Letter, we report the discovery of TNPSC in a 2D superconducting system with AFM order. We first introduce the theoretical description of such 2D TNPSC and then present its experimental realization in an AFM monolayer on top of an *s*-wave superconductor. The observed topological phase is established through the interplay of the pairing interaction of the substrate, the antiferromagnetic order of the magnetic monolayer, and the spin-orbit coupling (SOC) at the interface, which makes our results very general and not limited to a specific material system.

To investigate the properties of an AFM-MSH, we model a single layer of antiferromagnetically ordered magnetic adatoms with the symmetry of the (110) surface of a body-centered cubic (bcc) crystal. In such an AFM monolayer three different types of straight edges can form, as shown in Fig. 1a, which we name ferromagnetic (FM), zigzag (ZZ), and antiferromagnetic (AFM) edge in the following (the colors of the magnetic atoms in Fig. 1a indicate their spin orientations). We describe the electronic structure using a tight-binding model which reflects the presence of *s*-wave superconductivity, a Rashba spin-orbit interaction, and an AFM order of the magnetic moments, which interact with the electronic degrees of freedom via a magnetic exchange coupling [2] (see methods for additional details). The resulting superconducting band structure of the system, shown in Fig. 1b, exhibits 8 nodal points (NP) located on the boundary of the magnetic Brillouin zone (BZ). The curved arrow at each NP indicates its topological charge, with anticlockwise arrows for winding number of +1 and clockwise arrows for -1 [27,28] (see section S2 of the supplemental material for details). Pairs of cones from neighboring NPs are connected via a van-Hove point, which we refer to as β modes (see section S1 of the supplemental material for more details).

To model the three different types of edges, we consider semi-infinite systems [29] and present the corresponding edge spectral functions in Figs. 1c-e as a function of energy and momentum. The location of the bulk NPs, projected onto the momenta parallel to the different edges, are indicated by colored spheres. In addition to the bulk states, the FM and ZZ edges exhibit low energy modes that connect two NPs with opposite topological charge, which we refer to as $\alpha$ modes. In the local density of states (LDOS) for the FM and ZZ edges, shown in Figs. 1f,g, these α modes give rise to strong low-energy peaks., We note that the weak dispersion of the $\alpha$ mode at the FM edge results in a splitting of the low-energy peak. Moreover, the projection of the electronic structure onto the FM edge, combined with the emergence of trivial edge states near the bulk gap, leads to a splitting of the superconducting coherence peaks. In contrast, at the AFM edge, no low energy mode is present, as shown in Fig. 1e,h, while only a peak arising from the β mode is visible in the LDOS. Although β is a bulk mode, it is much less prominent in the FM and ZZ edge. These distinct and qualitatively different features in the LDOS between the FM and ZZ edges on one hand, and the AFM edge on the other, are a characteristic and unique signature of the topological electronic structure of the system.

An experimental system with an AFM layer on a bcc(110) superconductor, as used in the model for AFM-TNPSC, is realized by a pseudomorphic Mn monolayer on Nb(110), which was found to exhibit $c$(2x2) AFM order [30]. As seen in Fig. 2a, Mn forms islands that are elongated along [001] directions. In addition to long [001] edges, island edges along [1-10] and [1-11] are observed, which are the different edge types discussed in Fig. 1, namely the FM, the ZZ, and the AFM edge, respectively (see Fig 2b).

Figure 2c shows point spectra of the differential tunneling conductance (d$I$/d$U$) measured with a superconducting Nb-coated Cr tip at the positions indicated in Fig. 2a. Spectra taken on the clean Nb(110) (blue) show two coherence peaks at ±2.4 mV. With a Nb gap of 1.5 meV the tip gap amounts to $\Delta_{tip}$ = 0.9 meV. To be able to directly compare the experimental data with the calculated LDOS from our model in Fig. 1, we deconvolute all spectra and display them in Fig. 2d (see section S3 of the supplemental material). This results in an LDOS for a clean Nb surface which has symmetric coherence peaks at ±1.5 meV validating the deconvolution procedure. We attribute the non-zero intensity within the superconducting gap of the bare Nb LDOS to our tip, which is not a bulk superconductor. The deconvoluted LDOS obtained in the middle of the Mn island (green) exhibits a prominent peak at 0.8 meV, which we ascribe to the $\beta$ mode, while its negative energy counterpart ($\beta^-$) has a very low intensity. This resembles the characteristic LDOS for the Mn monolayer we reported earlier [30].

The LDOS of the three edges differ significantly: while the LDOS at the FM and the ZZ edges exhibits pronounced low-energy peaks, the LDOS at the AFM edge possesses only a peak that is located near the $\beta^+$ peak of the Mn monolayer. These features are in good agreement with our theoretical results shown in Fig. 1f-h obtained for infinitely long edges and can thus be considered as characteristic spectroscopic signatures of the AFM-TNPSC state. There are, however, some small differences between the experimental and theoretical results. First, for the FM edge, for example, our theoretical results reveal a peak split symmetrically around the Fermi energy, $\alpha^+$ and $\alpha^-$, whereas the experimental LDOS (Fig. 2d) exhibits one broad peak in the middle of the gap. However, a closer inspection of the experimental raw data (Fig. 2c) indeed reveals the presence of two features close to Fermi level: one split between positive and negative side of the tip gap and one at around 1.2 mV; these are not as obvious after the deconvolution procedure. In addition the theory predicts a splitting of the coherence peaks at the FM edge, whereas in the experimental data there is only one clear peak on either side of the gap. We propose that the intensity of some of the expected features is too small to resolve them. Second, at the ZZ edge, our theoretical results show a peak centered at zero energy, while the experimental observed peak in the LDOS is located at 90 µeV. Whether these small differences arises from the finite size of the experimentally investigated island (in contrast to the infinitely long edges considered theoretically) is presently unclear.

To investigate the spatial distribution of the spectroscopic fingerprints of our AFM-TNPSC we choose a Mn island with a particularly long FM edge, as shown in Fig. 3a. The insets show a spin-resolved constant-current image and a corresponding sketch of the antiferromagnetic order. Figure 3b shows d$I$/d$U$ maps related to $\alpha^\pm$ and $\beta^\pm$. We find that the $\beta^\pm$ state possesses the largest intensity in the interior of the Mn island, as expected for a bulk state, and in agreement with the spectra displayed in Fig. 2. In contrast, the $\alpha^\pm$ state possesses its largest intensity along the FM edge, exhibits a weaker intensity along the ZZ edge, and is essentially absent along the AFM edge, as expected from the dI/dU spectra shown in Fig.2. We note that the $\alpha^{\pm}$ state also exhibits a weak spatially oscillating intensity in the interior of the island, which we attribute to confinement effects. These spatial intensity maps clearly reveal the edge and bulk character of the $\alpha^{\pm}$ and $\beta^{\pm}$ states, respectively. Our experimental findings are in very good agreement with the theoretically computed maps shown in Fig. 3c, which was calculated for the same island shape and size as the experimental one, providing further support for the existence of the AFM-TNPSC state.

The localization of the $\alpha^{\pm}$ state at the FM edge is also clearly visible in the d*I*/d*U* spectra taken along the green arrow in Fig. 3a, presented as a waterfall plot in Fig. 3d. Inside the Mn island the $\beta^{+}$ state is dominant, while the $\alpha^{\pm}$ state possesses its largest spectral weight only at the island's edge. To quantify the localization we deconvolute the spectra, see Fig. 3e, and plot the resulting zero-bias intensity as a function of distance to the edge in Fig. 3f. The observed intensity decays exponentially on both sides of the edge, with a decay length of 1.5 nm towards the interior of the island, and of 1.0 nm on the Nb surface. The theoretically computed spatial dependence of the zero-bias intensity near the edge shows a very similar behavior, albeit with additional short-period oscillations on the Mn side.

Due to the finite size of the island, we also observe a spatial modulation of the intensity of the $\alpha^{\pm}$ state along the edges. To visualize this modulation, spectra were taken along the straight bottom FM edge between the positions of structural imperfections, see blue arrow in Fig. 3a, and presented as waterfall plot in Fig. 3g. These plots reveal that the intensities of the $\alpha^{-}$ or $\alpha^{+}$ branches are out-of-phase, exhibiting spatially alternating maxima. The corresponding deconvoluted data in Fig. 3h shows how the spectral weight of the $\alpha^{\pm}$ branches shifts across zero-bias along this particular FM edge. The different spatial structure of the $\alpha^{\pm}$ branches suggest that these branches are located at symmetric, non-zero energies. This, in turn, provides further evidence that the single peak in the experimentally obtained LDOS at the FM edge (see Fig. 2d) actually consists of two unresolved peaks that are located at symmetric energies around Fermi energy, in agreement with the theoretical results shown in Fig. 1f. Even though this modulation of the $\alpha^{\pm}$ intensity along the FM edge is not captured by the model, the spatial distribution of the edge and bulk states is overall in very good agreement with the characteristic properties of AFM-TNPSC as predicted by the model.

In conclusion, we developed a general model of a TNPSC within a 2D-AFM/ *s*-wave superconductor hybrid system, which is a realization of a time reversal invariant topological superconductor. We experimentally characterized this emergent quantum phase in a real material system, confirming the nodal-point superconducting state by observing the different edge modes as predicted by our model. Due to the rising interest in AFM and superconducting systems, we expect our findings to trigger new experimental and theoretical research in superconducting spintronics and quantum materials.

**Acknowledgements**

K.v.B., M.B. and R.L.C. acknowledge financial support from the Deutsche Forschungsgemeinschaft (DFG, German Research Foundation) Grants No. 418425860 and 459025680. M.B. acknowledges the Polish Ministry of Education and Science within Project No. 0512/SBAD/2220 realized at Faculty of Materials Engineering and Technical Physics, Poznan University of Technology. R.W. acknowledges financial support by the European Union via the ERC Advanced Grant ADMIRE. E.M. and R.W. gratefully acknowledge funding by the Cluster of Excellence 'Advanced Imaging of Matter' (EXC 2056 - project ID 390715994) of the Deutsche Forschungsgemeinschaft (DFG). E.M. acknowledges financial support from the Australian Research Council under project DP200101118. D.K.M. acknowledges support by the U. S. Department of

Energy, Office of Science, Basic Energy Sciences, under Award No. DE-FG02-05ER46225. The authors are thankful to Dr. André Kubetzka for his support in the initial establishment of the experiment.

**Authors Contribution**

M.B., R.L.C., and E.M. contributed equally to this work. M.B. and R.L.C. conceived and executed the experiments and analyzed the experimental data. K.v.B. supported the experiment. E.M. developed the TB model, performed the calculations, and analyzed the model data. D.K.M. validated the TB model. M.B., R.L.C., and E.M. drafted the manuscript. M.B., R.L.C., E.M., D.K.M., and K.v.B. wrote the manuscript. All authors interpreted the data, discussed the results, and commented on the manuscript.

**Methods**

Monolayer thick Mn islands were grown on top of an unreconstructed Nb(110) single crystal via physical vapor deposition from a crucible under UHV conditions with a base pressure of about $1.0 \times 10^{-10}$ mbar following the procedure as described by Lo Conte *et al.* [30]. Before the Mn deposition, the Nb(110) substrate was cleaned by a series of 50 sec-long flashes in UHV with base pressure of $1.0 \times 10^{-10}$ mbar, during which a maximum temperature of about 2400°C is reached. Samples after growth were *in situ* transferred to a home-built LT STM system operated at 1.3 K in UHV.

All STM experiments were caried out using a Nb-coated superconducting tip. The tip was made by indentation of a bulk Cr tip into clean Nb prior to the experiments. The *dI/dU* measurements were performed using a lock-in technique by adding a small modulation voltage with a frequency of 4333 Hz to the bias voltage. *dI/dU* maps were recorded in a constant-height mode after stabilization with the tip parked in the middle of the island. The effective electronic temperature of the tunneling junction was 1.8 K as determined by a BCS fit to a *dI/dU* spectrum of superconducting Nb(110) acquired with a normal tip. Details of the numerical deconvolution are available in the supplement material. All data has been processed using a self-written python code.

The tight-binding model used is given by

$$H = t_0 \sum_{\langle \mathbf{r},\mathbf{r}'\rangle,\sigma} c_{\mathbf{r}\sigma}^\dagger c_{\mathbf{r}'\sigma} + t_1 \sum_{\langle\langle \mathbf{r},\mathbf{r}'\rangle\rangle,\sigma} c_{\mathbf{r}\sigma}^\dagger c_{\mathbf{r}'\sigma} + t_2 \sum_{\langle\langle\langle \mathbf{r},\mathbf{r}'\rangle\rangle\rangle,\sigma} c_{\mathbf{r}\sigma}^\dagger c_{\mathbf{r}'\sigma} - \mu \sum_{\mathbf{r}} c_{\mathbf{r}\sigma}^\dagger c_{\mathbf{r}\sigma}$$

$$-i\alpha \sum_{\langle \mathbf{r},\mathbf{r}'\rangle,\langle\langle \mathbf{r},\mathbf{r}'\rangle\rangle,\alpha,\beta} c_{\mathbf{r}\alpha}^\dagger \left(\boldsymbol{\sigma}_{\alpha\beta} \times \frac{\mathbf{r}'-\mathbf{r}}{|\mathbf{r}'-\mathbf{r}|}\right)_z c_{\mathbf{r}\beta} + J \sum_{\mathbf{r}} e^{i\mathbf{Q}\cdot\mathbf{r}}(c_{\mathbf{r}\uparrow}^\dagger c_{\mathbf{r}\uparrow} - c_{\mathbf{r}\downarrow}^\dagger c_{\mathbf{r}\downarrow}) + \Delta \sum_{\mathbf{r}}(c_{\mathbf{r}\uparrow}^\dagger c_{\mathbf{r}\downarrow}^\dagger + c_{\mathbf{r}\downarrow} c_{\mathbf{r}\uparrow}), \quad (1)$$

where $t_0, t_1, t_2$ are the nearest, next-nearest, and next-next-nearest neighbor hopping parameters, respectively; $\mu$ is the chemical potential; $\alpha$ is the Rashba spin-orbit coupling; $J$ is the exchange coupling; and $\Delta$ is the superconducting order parameter. The hopping directions are $\left(\pm\frac{a}{2}, \pm\frac{b}{2}\right)$ for nearest neighbor, $(\pm a, 0)$ for next-nearest neighbor, and $(0, \pm b)$ for next-next-nearest neighbor. The ordering wavevector, **Q**, is $\left(\frac{2\pi}{a}, 0\right)$ or equivalently $\left(0, \frac{2\pi}{b}\right)$. The parameters used in the main text are $(t_0, t_1, t_2, \mu, \alpha, J, \Delta) = (1.1, 0.9, 1, 3.5, 0.5, 3.1, 1.5) \times 1.7$ meV. For an infinite system, Eq. (1) can be expressed in momentum space as

$$H = \sum_{\mathbf{k}} \psi_{\mathbf{k}}^\dagger \{\eta_0 \tau_z(\varepsilon_{\mathbf{k}}\sigma_0 + \beta_{\mathbf{k}}\sigma_y) + \eta_x \tau_z(f_{\mathbf{k}}\sigma_0 + \boldsymbol{\alpha}_{\mathbf{k}} \cdot \boldsymbol{\sigma}) + J\eta_z \tau_0 \sigma_z + \Delta \eta_0 \tau_x \sigma_0\} \psi_{\mathbf{k}} \quad (2)$$

$$\varepsilon_{\mathbf{k}} = 2t_1 \cos(k_x a) + 2t_2 \cos(k_y b) - \mu \quad (2.1)$$

$$f_{\mathbf{k}} = 4t_0 \cos\left(\frac{k_x a}{2}\right) \cos\left(\frac{k_y b}{2}\right) \quad (2.2)$$

$$\boldsymbol{\alpha}_{\mathbf{k}} = -4\alpha \left(-\sqrt{\frac{2}{3}} \cos\left(\frac{k_x a}{2}\right) \sin\left(\frac{k_y b}{2}\right), \quad \sqrt{\frac{1}{3}} \sin\left(\frac{k_x a}{2}\right) \cos\left(\frac{k_y b}{2}\right), \quad 0\right) \quad (2.3)$$

$$\beta_{\mathbf{k}} = -2\alpha \sin(k_x a) \quad (2.4)$$

where the spinor is given by

$$\psi_{\mathbf{k}}^\dagger = \begin{pmatrix} a_{\mathbf{k}\uparrow}^\dagger & a_{\mathbf{k}\downarrow}^\dagger & a_{-\mathbf{k}\downarrow} & -a_{-\mathbf{k}\uparrow} & b_{\mathbf{k}\uparrow}^\dagger & b_{\mathbf{k}\downarrow}^\dagger & b_{-\mathbf{k}\downarrow} & -b_{-\mathbf{k}\uparrow} \end{pmatrix}. \quad (3)$$

$a_{\mathbf{k}\sigma}$ and $b_{\mathbf{k}\sigma}$ are the annihilation operators for A (up moments) and B (down moments) sublattices, respectively; $\eta_a$, $\tau_a$, and $\sigma_a$ are Pauli matrices in sublattice, particle-hole, and spin space, respectively.

* These authors contributed equally.
§ Corresponding Authors
mbazarni@physnet.uni-hamburg.de; rolocont@physnet.uni-hamburg.de; eric.mascot@unimelb.edu.au

**References**

[1]  Nadj-Perge, S., Drozdov, I. K., Bernevig, B. A., and Yazdani, A. Proposal for realizing Majorana fermions in chains of magnetic atoms on a superconductor. *Phys. Rev. B* **88**, 020407(R) (2013).

[2]  Li, J., Neupert, T., Wang, Z., MacDonald, A.H., Yazdani, A. and Andrei Bernevig B. Two-dimensional chiral topological superconductivity in Shiba lattices. *Nat. Commun.* **7**, 12297 (2016).

[3]  Bedow, J., Mascot, E., Posske, T., Uhrig, G. S., Wiesendanger, R., Rachel, S., and Morr, D. K. Topological superconductivity induced by a triple-q magnetic structure *Phys. Rev. B* **102**, 180504 (2020).

[4]  Mascot, E., Bedow, J., Graham, M., Rachel, S., and Morr, D. K.Topological superconductivity in skyrmion lattices. *npj Quantum Mater.* **6**, 6 (2021).

[5]  Steffensen, D., Christensen, M. H., Andersen, B. M., and Kotetes, P. Topological superconductivity induced by magnetic texture crystals. *Phys. Rev. Research* **4**, 013225 (2022).

[6]  Nadj-Perge, S., Drozdov, I.K., Li, J., Chen, H., Jeon, S., Seo, J., MacDonald, A.H., Bernevig, B.A. and Yazdani, A. Observation of Majorana fermions in ferromagnetic atomic chains on a superconductor. Science, **346**, 602-607 (2014).

[7]  Feldman, B.E., Randeria, M.T., Li, J., Jeon, S., Xie, Y., Wang, Z., Drozdov, I.K., Andrei Bernevig, B. and Yazdani, A. High-resolution studies of the Majorana atomic chain platform. *Nat. Phys.*, **13**, 286-291 (2017).

[8]  Schneider, L., Beck, P., Posske, T., Crawford, D., Mascot, E., Rachel, S., Wiesendanger, R. and Wiebe, J. Topological Shiba bands in artificial spin chains on superconductors. *Nature Phys.*, **17**, 943-948 (2021).

[9]  Ménard, G.C., Guissart, S., Brun, C., Leriche, R.T., Trif, M., Debontridder, F., Demaille, D., Roditchev, D., Simon, P. and Cren, T. Two-dimensional topological superconductivity in Pb/Co/Si (111). *Nat. Commun.*, **8**, 1-7 (2017).

[10] Palacio-Morales, A., Mascot, E., Cocklin, S., Kim, H., Rachel, S., Morr, D.K. and Wiesendanger, R.

Atomic-scale interface engineering of Majorana edge modes in a 2D magnet-superconductor hybrid system. *Sci. Adv.* **5**: eaav6600 (2019).

[11] Kezilebieke, S., Huda, M.N., Vaňo, V., Aapro, M., Ganguli, S.C., Silveira, O.J., Głodzik, S., Foster, A.S., Ojanen, T. and Liljeroth, P. Topological superconductivity in a van der Waals heterostructure. *Nature* **588**, 424 (2020).

[12] Kezilebieke, S., Vano, V., Huda, M.N., Aapro, M., Ganguli, S.C., Liljeroth, P. and Lado, J.L. Moiré-enabled topological superconductivity. *Nano Lett.* **22**, 328 (2022).

[13] Nayak, A.K., Steinbok, A., Roet, Y., Koo, J., Margalit, G., Feldman, I., Almoalem, A., Kanigel, A., Fiete, G.A., Yan, B. and Oreg, Y. Evidence of topological boundary modes with topological nodal-point superconductivity. *Nat. Phys.* **17**, 1413 (2021).

[14] Sato, M. Nodal structure of superconductors with time-reversal invariance and $Z_2$ topological number. *Phys. Rev. B*, **73**, 214502 (2006).

[15] Béri, B. Topologically stable gapless phases of time-reversal-invariant superconductors. *Phys. Rev. B* **81**, 134515 (2010).

[16] Prada, E., San-Jose, P., de Moor, M.W., Geresdi, A., Lee, E.J., Klinovaja, J., Loss, D., Nygård, J., Aguado, R. and Kouwenhoven, L.P. From Andreev to Majorana bound states in hybrid superconductor–semiconductor nanowires. *Nat. Rev. Phys.* **2**, 575-594 (2020).

[17] Beenakker C. Search for Majorana Fermions in Superconductors. *Annual Review Condensed Matter Physics* **4**, 113 (2013).

[18] Schnyder, A.P., Ryu, S., Furusaki, A. and Ludwig, A.W. Classification of topological insulators and superconductors in three spatial dimensions. *Phys. Rev. B*. **78**, 195125 (2008).

[19] Chiu, C.K., Teo, J.C., Schnyder, A.P. and Ryu, S. Classification of topological quantum matter with symmetries. *Rev. Mod. Phys.*, **88**, 035005 (2016).

[20] Schnyder, A.P. and Brydon, P.M. Topological surface states in nodal superconductors. *J. Phys: Cond. Mat.* **27**, 243201 (2015).

[21] Baum, Y., Posske, T., Fulga, I.C., Trauzettel, B. and Stern, A. Gapless topological superconductors: Model Hamiltonian and realization. *Phys. Rev. B* **92**, 045128 (2015).

[22] Kao, J.T., Huang, S.M., Mou, C.Y. and Tsuei, C.C. Tunneling spectroscopy and Majorana modes emergent from topological gapless phases in high-$T_c$ cuprate superconductors. *Phys. Rev. B* **91**, 134501 (2015).

[23] Zhu, G.Y., Wang, Z. and Zhang, G.M. Two-dimensional topological superconducting phases emerged from *d*-wave superconductors in proximity to antiferromagnets. *Europhys. Lett.* **118**, 37004 (2017).

[24] Bouhon, A., Schmidt, J. and Black-Schaffer, A.M. Topological nodal superconducting phases and topological phase transition in the hyperhoneycomb lattice. *Phys. Rev. B* **97**, 104508 (2018).

[25] Brzezicki, W. and Cuoco, M. Nodal *s*-wave superconductivity in antiferromagnetic semimetals. *Phys. Rev. B* **97**, 064513 (2018).

[26] Lu, M.L., Wang, Y., Zhang, H.Z., Chen, H.L., Cui, T.Y. and Luo, X. Chiral symmetry protected topological nodal superconducting phase and Majorana Fermi arc. *Chinese Physics B*, (2022). https://doi.org/10.1088/1674-1056/ac7208.

[27] Béri, B. Topologically stable gapless phases of time-reversal-invariant superconductors, *Phys. Rev. B.* **81**, 134515 (2010).

[28] Margalit, G., Berg, E. and Oreg, Y. Theory of multi-orbital topological superconductivity in transition metal dichalcogenides. *Annals of Physics* **435**, 168561 (2021).

[29] Lopez Sancho, M.P., Lopez Sancho, J.M., Sancho, J. M. L. and Rubio, J. Highly convergent schemes for the calculation of bulk and surface Green functions. *J. Phys. F: Met. Phys.* **15,** 851 (1985)

[30] Lo Conte, R., Bazarnik, M., Palotás, K., Rózsa, L., Szunyogh, L., Kubetzka, A., von Bergmann, K., and Wiesendange, R. Coexistence of antiferromagnetism and superconductivity in Mn/Nb (110). *Phys. Rev. B* **105**, L100406 (2022).

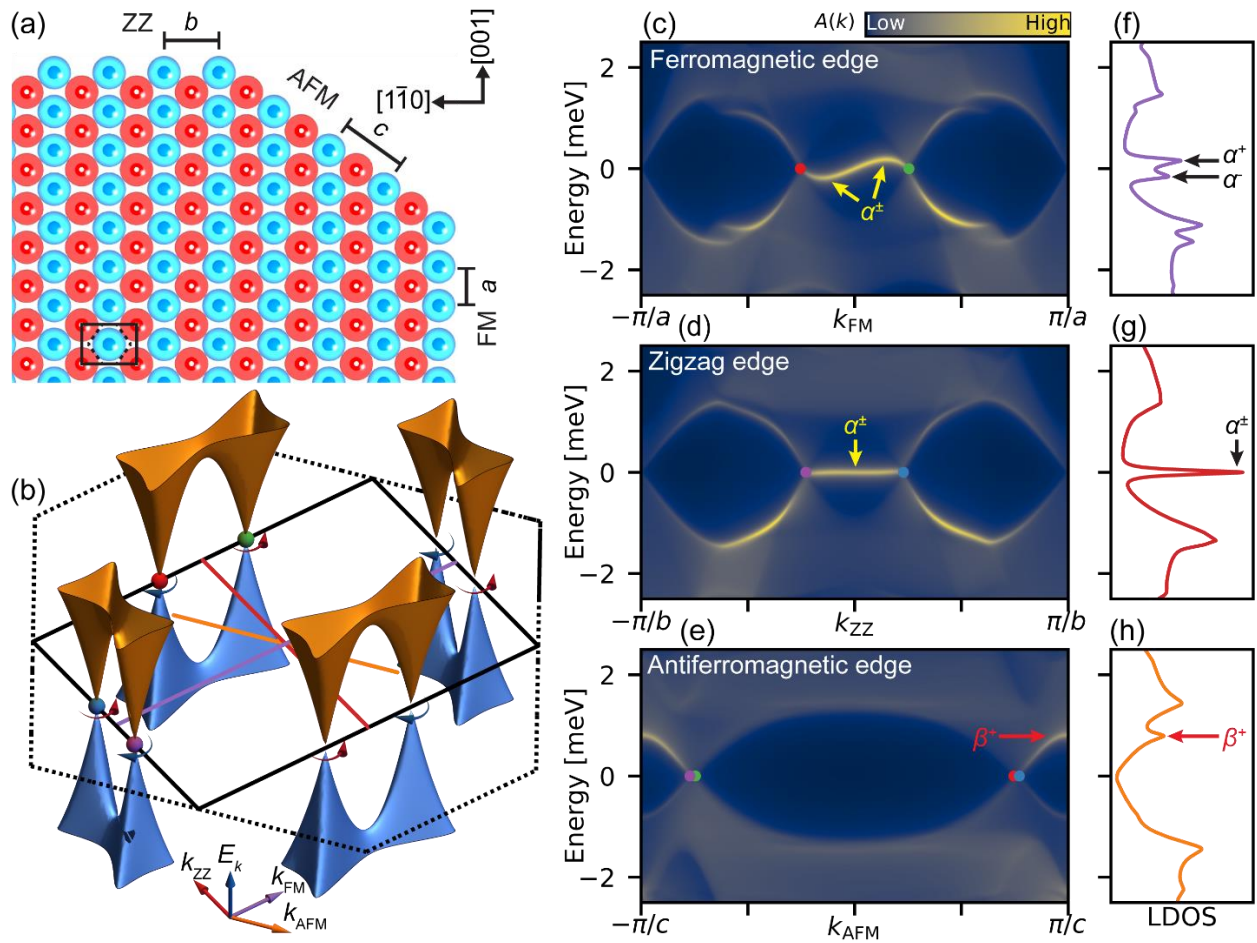

Figure 1. Electronic properties of an AFM-MSH. (a) Real space structure of an antiferromagnetically ordered mono-atomic layer with the symmetry of the (110) surface of a body-centered cubic crystal. (b) Superconducting quasiparticle dispersion. The dispersion shows nodal points on the boundary of the magnetic Brillouin zone, shown as a black rectangle (the structural Brillouin zone is shown as dashed lines). The winding number associated with each nodal point is indicated by the curved arrows (+1 red, -1 blue). (c-e) Spectral function for each edge type and (f-h) corresponding LDOS at (c,f) a FM edge, (d,g) a ZZ edge, and (e,h) an AFM edge. In addition to the edge modes, the spectral functions have contributions from the 2D bulk electronic structure, which originate from the projected bulk electronic band structure shown in (b) onto the respective momentum axes, as indicated by the colored lines in (b) (see supplementary Fig. S2 for individual projections). A dispersive edge band connects the projected nodal points at the FM edge and a flat edge band for the ZZ edge (yellow arrows), resulting in strong low-energy peaks in the LDOS (black arrows). For the AFM edge, a state corresponding to the van-Hove singularity connecting two neighboring cones appears and is indicated by red arrows. Parameters for the model are given in the methods section.

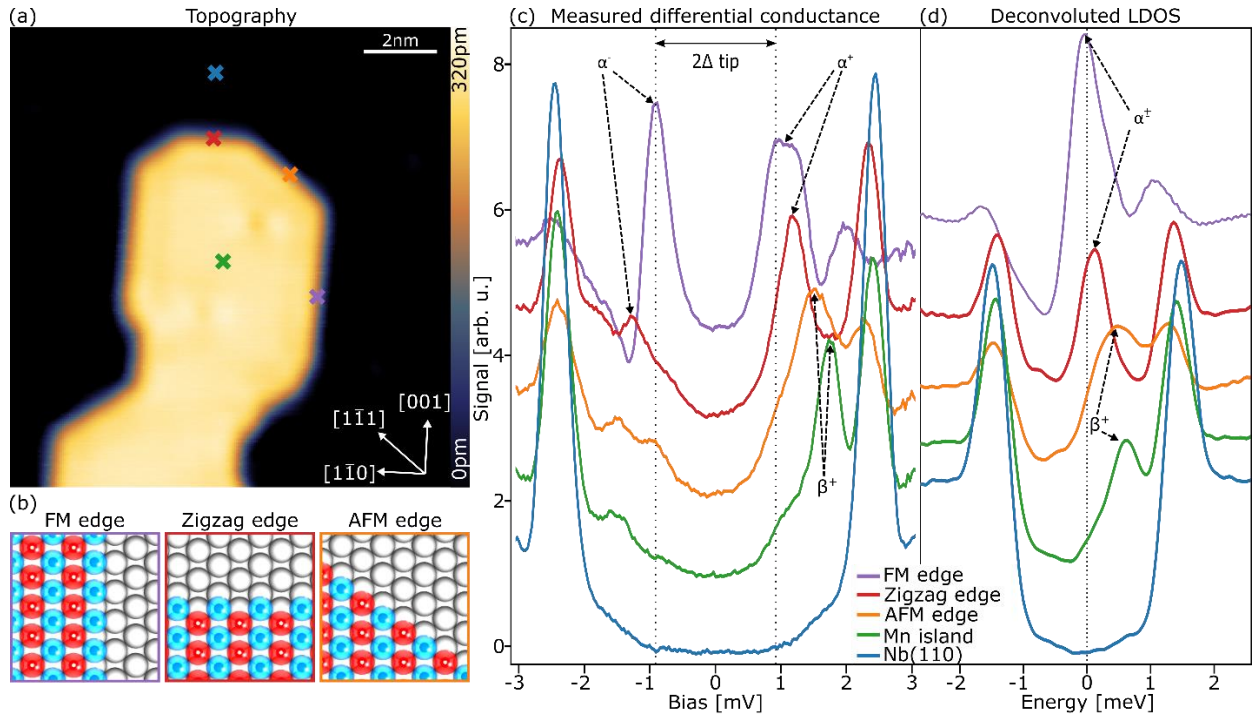

Figure 2. Local tunneling spectroscopy on a Mn/Nb(110) island. (a) Topography image of a single Mn peninsula with well-developed edges along the main crystallographic directions. (b) Schematics of the atomic and spin structure of a FM, ZZ, and AFM edge, respectively. (c) Tunneling point spectra acquired with a superconducting Nb-coated Cr tip at the positions indicated in panel (a). (d) Deconvoluted LDOS obtained from the spectra displayed in panel (c) [cf. supp. material for deconvolution details]. Tunneling parameters: (a) tunneling current $I_t$ = 1 nA, bias $U$ = 20 mV; (c) stabilization tunneling current $I_{ts}$ = 1 nA, stabilization bias $U_{stab}$ = 20 mV, modulation bias $U_{mod}$ = 100 µV, vertical tip shift towards the sample's surface after stabilization $Z_{approach}$ = 80 pm.

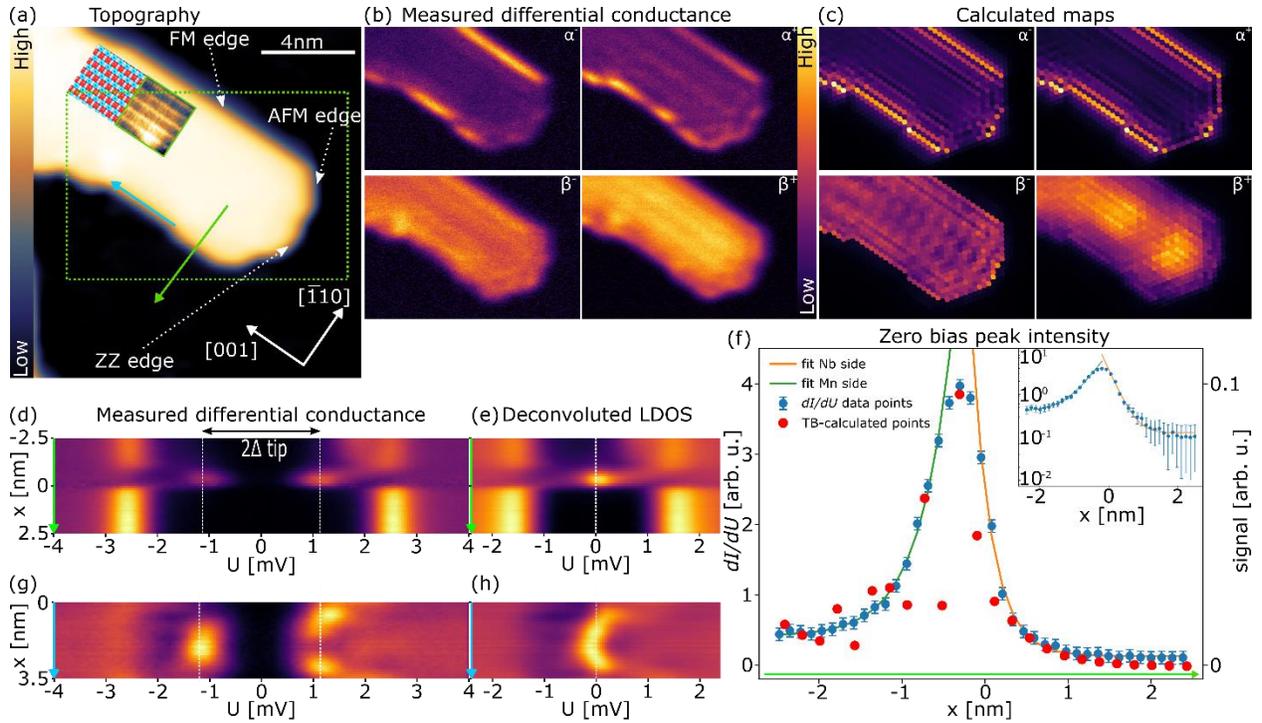

Figure 3. Spatial dependence of spectroscopic features. (a) A topography image of a defect-free Mn island. The insets show an SP-STM image of part of the island and a sketch of the *c*(2x2) antiferromagnetic state. The green rectangle indicates the sample area of panel (b). (b) Differential tunneling conductance maps acquired with a superconducting Nb-coated Cr tip at characteristic energies as indicated for $\alpha^{\pm}$ and $\beta^{\pm}$. (c) Tight-binding model calculated maps for the same spectroscopic features as in panel (b). (d) Spectra along the line marked with a green arrow in panel (a) crossing one of the FM edges. (e) Deconvoluted LDOS obtained from the data in panel (d). (f) Intensity of the zero-bias peak as a function of the distance from the edge. The green and orange curves are exponential decay fits to experimental data towards the Mn and Nb, respectively. The inset shows a semi-log plot for experimental data and fits. (g) Spectra along the line marked with a blue arrow in panel (a) along one of the ferromagnetic edges. (h) Deconvoluted LDOS from panel (g). Through the panels (d-f) the 0 on the x-axis is set to the half-height of the Mn island in the topography profile along the green arrow. Tunneling parameters: (a) $I_t$ = 1 nA, $U$ = 50 mV; (b) $I_{ts}$ = 1 nA, $U_{stab}$ = 20 mV, $U_{mod}$ = 100 µV, $Z_{approach}$ = 80 pm, for $\alpha^{\pm}$ $U$ = ±1.05 mV, for $\beta^{\pm}$ $U$ = ±1.9 mV; (d) $I_{ts}$ = 1 nA, $U_{stab}$ = 20 mV, $U_{mod}$ = 100 µV, $Z_{approach}$ = 100 pm; (g) $I_{ts}$ = 1 nA, $U_{stab}$ = 20 mV, $U_{mod}$ = 100 µV, $Z_{approach}$ = 100 pm.

# SUPPLEMENTAL MATERIAL

## Antiferromagnetism-driven two-dimensional topological nodal-point superconductivity

Maciej Bazarnik[1,2,*,§], Roberto Lo Conte[1,*,§], Eric Mascot[1,3,*,§], Dirk K. Morr[4], Kirsten von Bergmann[1], Roland Wiesendanger[1]

[1]*Department of Physics, University of Hamburg, D-20355 Hamburg, Germany*
[2]*Institute of Physics, Poznan University of Technology, Piotrowo 3, 60-965 Poznan, Poland*
[3]*School of Physics, University of Melbourne, Parkville, VIC 3010, Australia*
[4]*Department of Physics, University of Illinois at Chicago, Chicago, IL 60607*

**\*** These authors contributed equally.
§ Corresponding Authors
 mbazarni@physnet.uni-hamburg.de; rolocont@physnet.uni-hamburg.de; eric.mascot@unimelb.edu.au

**Content:**

**Section S1. The calculated band structure and density of states**

**Section S2. Topology**

**Section S3. Numerical deconvolution of superconductor-vacuum-superconductor spectra**

## S1. The calculated band structure and density of states

The calculated band structure and density of states are shown in Fig. S1. The band structure shows van-Hove singularities between neighboring NPs at $\left(\frac{\pi}{a},0\right)$, $\left(0,\frac{\pi}{b}\right)$, and between non-neighboring NPs near $\left(\frac{0.63\pi}{a},\frac{0.57\pi}{b}\right)$. In Fig. S2, we present the projections of the bulk band structure onto the momentum axes parallel to each edge type. We also show the band structure for a cylinder geometry, where the system is finite in the direction perpendicular to the edge and periodic parallel to the edge. The cylinder geometry shows a similar band structure to the projected bulk band structure with the addition of edge bands.

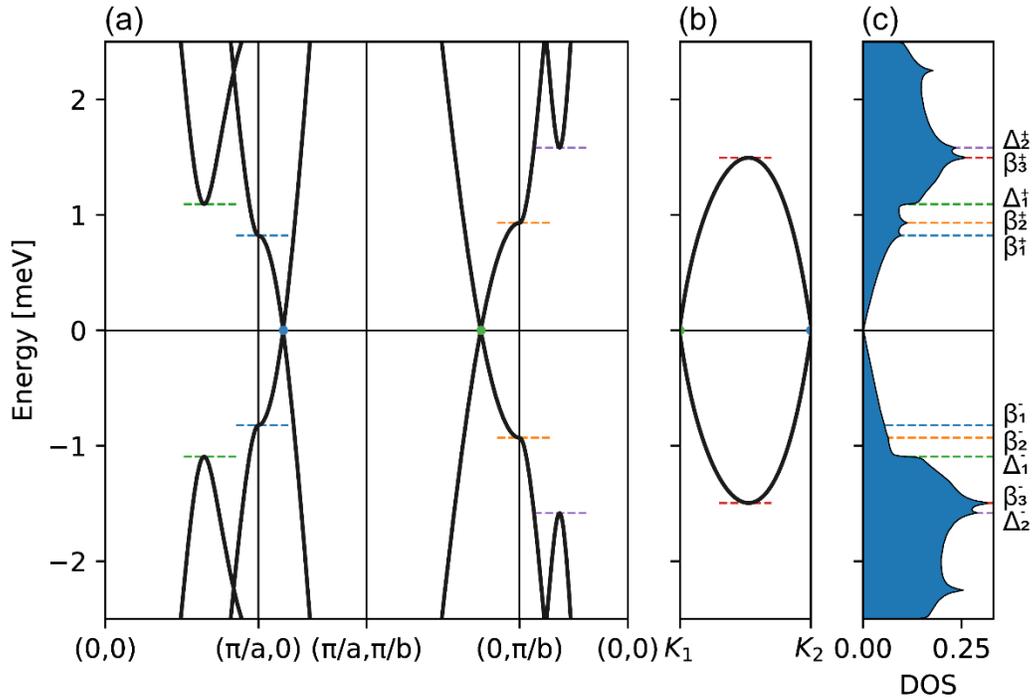

Figure S1. Band structure and density of states. (a) Band structure along a high-symmetry path. The blue and orange dashed lines indicate van-Hove singularities and the green and purple dashed lines indicate the coherence peaks. (b) Band structure between two non-neighboring NPs. Dashed red line indicates a van-Hove singularity. (c) Density of states with corresponding van-Hove singularities and coherence peaks indicated by dashed lines.

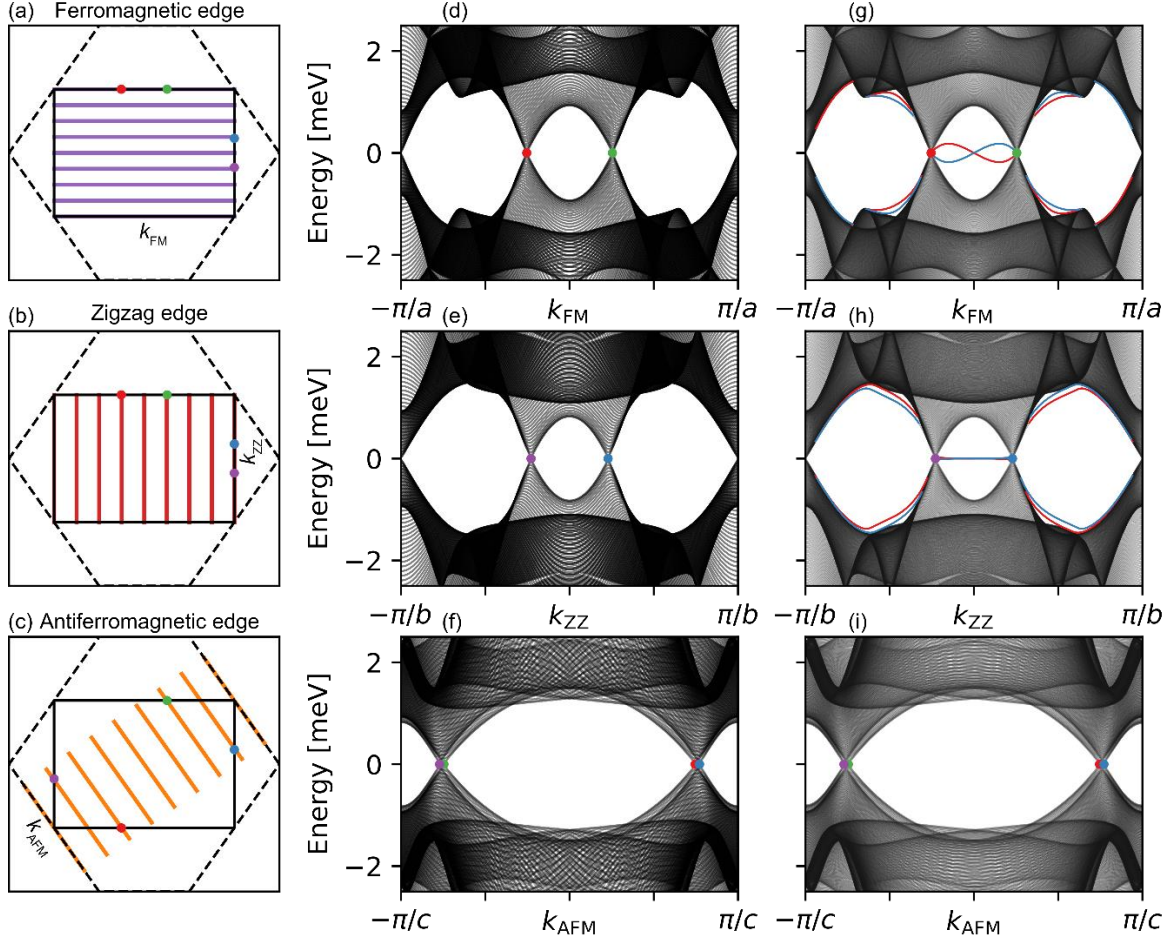

Figure S2. Band structure projections. (a-c) Momentum axes parallel to (a) FM, (b) ZZ, and (c) AFM edges. The colored lines indicate the momenta in (d-e). The colored dots indicate the locations of the NPs. The black rectangle shows the magnetic BZ and the dashed rectangle shows the structural BZ. (d-f) Bulk band structure projections for the corresponding edges. (g-i) Band structure for a cylinder geometry for the corresponding edges.

## S2. Topology

The Hamiltonian in Eq. (2) has the symmetries, $T = \eta_x \tau_0 \sigma_y K$, $C = \eta_0 \tau_y \sigma_y K$, and $S = \eta_x \tau_y \sigma_0$. The symmetries square to $T^2 = -1$ and $C^2 = S^2 = 1$, yielding symmetry class DIII [S1]. We then transform the Hamiltonian in Eq. (2) to the basis where $S$ is diagonal to get an off-diagonal matrix,

$$U^\dagger S U = \begin{pmatrix} 1 & \\ & -1 \end{pmatrix}, U^\dagger H_{\mathbf{k}} U = \begin{pmatrix} & h_{\mathbf{k}} \\ h_{\mathbf{k}}^\dagger & \end{pmatrix}. \tag{S1}$$

Next, we project the energies to $\pm 1$,

$$Q_{\mathbf{k}} = \sum_n |n_{\mathbf{k}}\rangle \text{sign}(E_{n,\mathbf{k}})\langle n_{\mathbf{k}}| = \begin{pmatrix} & q_{\mathbf{k}} \\ q_{\mathbf{k}}^\dagger & \end{pmatrix}, \tag{S2}$$

where $E_{n,\mathbf{k}}$ are the energies and $|n_{\mathbf{k}}\rangle$ are the eigenvectors of the Hamiltonian in Eq. (S1). In Fig. S3, we show the characteristic angle $e^{i\theta_{\mathbf{k}}} = \det(q_{\mathbf{k}})$. The winding of $\theta_{\mathbf{k}}$ around each nodal point defines its topological charge, which is expressed as

$$\nu = \frac{1}{2\pi i} \oint d\mathbf{k} \cdot \text{Tr}[q_{\mathbf{k}}^{-1} \nabla_{\mathbf{k}} q_{\mathbf{k}}] = \frac{\Delta \theta_{\mathbf{k}}}{2\pi}. \tag{S3}$$

The winding number of the NPs alternate between +1 and -1 along the magnetic BZ boundary and NPs at **k** and -**k** have the same winding number.

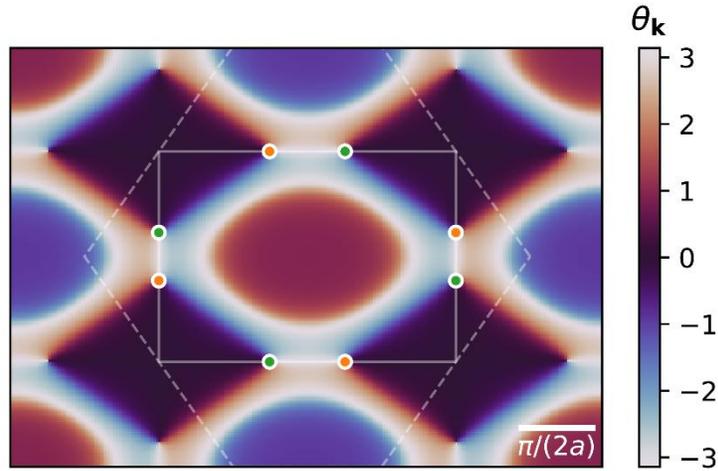

Figure S3. Characteristic angle. The magnetic BZ is indicated by solid lines and the structural BZ is indicated by dashed lines. NPs with winding number +1 are marked by green points and NPs with winding number -1 are marked by orange points.

## S3. Numerical deconvolution of superconductor-vacuum-superconductor spectra

In our scanning tunneling spectroscopy experiments, we employed a superconducting Nb-coated Cr tip in order to enhance the experimental energy resolution. As a result, the acquired d$I$/d$U$ spectra do not directly correspond to the LDOS of the sample, as for the case of normal metallic tips, but rather to the convolution of the density of states of the sample and of the superconducting tip. Accordingly, in order to retrieve the actual LDOS of the sample we perform a numerical deconvolution of the measured d$I$/d$U$ spectra [S2,S3,S4]. The tunneling conductance can be expressed as:

$$\frac{dI}{dU}(U,T) \propto \int_{-\infty}^{+\infty} \rho_S(E) \frac{\partial \rho_T(E-eU)}{\partial U} [f(E-eU,T) - f(E,T)] dE$$

$$+ \int_{-\infty}^{+\infty} \rho_S(E) \rho_T(E-eU) \frac{\partial f(E-eU,T)}{\partial U} dE \quad (S4)$$

where $\rho_S(E)$ is the energy-dependent LDOS of the sample below the tip apex, $\rho_T(E)$ is the LDOS at the tip apex, $f(E,T)$ is the Fermi-Dirac distribution function, $T$ is the experimental temperature, and $U$ is the applied bias between tip and sample.

We discretize the integral in Eq. (S4) into $N$ points in the interval $U \in$ [-4, +4] meV. Treating the measured d$I$/d$U$ spectrum as a column vector [d$I$/d$U$] of length $N$, Eq. (S4) can be expressed as a multiplication of an (unknown) sample LDOS vector [$\rho_S$] of length $N$ with an $N \times N$ matrix $\mathbf{A}$:

$$\left[\frac{dI}{dU}\right] = \mathbf{A}[\rho_S]. \quad (S5)$$

Accordingly, the matrix elements of $\mathbf{A}$ are:

$$A_{ij} = \left(\frac{\partial \rho_T(E_j - eU_i)}{\partial U}[f(E_j - eU_i, T) - f(E_j, T)] + \rho_T(E_j - eU_i)\frac{\partial f(E_j - eU_i, T)}{\partial U}\right) \times \delta E, \quad (S6)$$

where $E_j$ are the discrete energy values and $U_i$ the bias voltage values of the acquired tunneling spectrum. Numerically computing a matrix inverse $\mathbf{A}^{-1}$ [S3,S4] and multiplying the result with [d$I$/d$U$] yields an approximate sample LDOS:

$$[\rho_S] \sim \mathbf{A}^{-1}\left[\frac{dI}{dU}\right]. \quad (S7)$$

The DOS of the superconducting Nb-coated Cr tip is assumed to be a BCS DOS with a phenomenological broadening parameter Γ as discussed by Dynes et al. [S5]:

$$\rho_T(E) = \rho_0 Re\left[\frac{E - i\Gamma}{\sqrt{(E - i\Gamma)^2 - \Delta_T^2}}\right], \quad (S8)$$

where $\rho_0$ is the normal conducting DOS of the tip, and *Re* indicates the real part.

Parameters used for the deconvolution are as follows: T = 1.8 K, for $\rho_T(E)$ we assume a superconducting DOS with a $\Delta_T = 0.95$ meV for measurements shown in Fig. 2 and $\Delta_T = 1.05$ meV for measurements shown in Fig. 3. To account for the thin Nb coating of the Cr tip the Γ parameter is set to 1e-4.

**References**

[S1] Chiu, C., Teo, J. C. Y., Schnyder, A. P., and Ryu, S. Classification of topological quantum matter with symmetries. *Rev. Mod. Phys.* **88**, 035005 (2016).

[S2] Pillet, J.-D., Quay, C. H. L., Morfin, P., Bena, C., Yeyati, A. L., and Joyez, P. Andreev bound states in supercurrent-carrying carbon nanotubes revealed. *Nat. Phys*. **6**, 965 (2010).

[S3] Choi, D.-J., Rubio-Verdú, C., de Bruijckere, J., Ugeda, M. M., Lorente, N., and Pascual, J. I. Mapping the orbital structure of impurity bound states in a superconductor. *Nat. Commun*. **8**, 15175 (2017).

[S4] Lo Conte, R., Bazarnik, M., Palotás, K., Rózsa, L., Szunyogh, L., Kubetzka, A., von Bergmann, K., and Wiesendange, R. Coexistence of antiferromagnetism and superconductivity in Mn/Nb (110). *Phys. Rev. B* **105**, L100406 (2022).

[S5] Dynes, R. C., Narayanamurti, V., and Garno, J. P. Direct measurement of quasiparticle-lifetime broadening in a strong-coupled superconductor. *Phys. Rev. Lett.* **41**, 1509 (1978).